\documentclass[useAMS,usenatbib]{mn2e}
\usepackage{graphicx,graphics,amsmath}
\title[Observations of 1A 1118-61 during an outburst]{{\em RXTE}-PCA observations of 1A 1118--61: timing and spectral studies during an outburst}

\author[Jincy Devasia, Marykutty James, Biswajit Paul and Kavila Indulekha]{Jincy Devasia$^{1,2}$\thanks{E-mail: jincydevasia@yahoo.com}, Marykutty James$^{1,2}$\thanks{E-mail:marykuttykjames@yahoo.co.in}, Biswajit Paul$^{2}$,  and Kavila Indulekha$^{1}$\\
$^{1}$School of Pure and Applied Physics, Mahatma Gandhi University, Kottayam-686560, Kerala, India\\
$^{2}$Raman Research Institute, Sadashivanagar, C. V. Raman Avenue, Bangalore 560080, India}
\begin{document}

\maketitle
 \label{firstpage}

\begin{abstract}

We report detailed timing and spectral analysis of {\em RXTE}-PCA data obtained from observations during the outburst of a transient X-ray pulsar 1A 1118--61 in January 2009. The pulse profile showed significant evolution during the outburst and also significant energy dependence $-$ a double peaked profile upto 10 keV and a single peak at higher energy. We have also detected quasi-periodic oscillations (QPO) at 0.07--0.09 Hz. The rms value of the QPO is 5.2 $\%$ and it shows a significant energy dependence with highest rms of 7$\%$ at 9 keV. The QPO frequency changed from 0.09 Hz to 0.07 Hz within 10 days. The magnetic field strength calculated using the QPO frequency and the X-ray luminosity is in agreement with the magnetic field strength measured from the energy of the cyclotron absorption feature detected in this source. The 3-30 keV energy spectrum over the 2009 outburst of 1A 1118--61 can be well fitted with a partial covering power-law model with a high energy cutoff and an iron fluorescence line emission. The pulse phase resolved spectral analysis shows that the partial covering and high energy cutoff model parameters have significant changes with the pulse phase.
 
\end{abstract}


\section{INTRODUCTION}

   The hard X-ray transient pulsar 1A 1118--61 was discovered with the Rotation Modulation Collimator ({\em RMC}) experiment on {\em Ariel V} in 1974 \citep{Eyles1975}. Pulsations were detected in this source  with a period of 405.3 s \citep{Ives1975} and the optical counterpart of this source was identified as  He 3--640 = WRA 793 \citep{Chevalier1975} which is a highly reddened Be star. The star has a visual magnitude of V = 12.1 and is classified as a O9.5IV-Ve \citep{Janot-Pacheco1981, Motch1988} with strong Balmer emission lines indicating the presence of an extended envelope. The UV spectrum shows many absorption features; especially the CIV line indicating a stellar outflow. The P-Cygni profile gives a wind velocity in the range of 1600 $\pm$ 300 km s$^{-1}$ and the general spectral profile is similar to that of the optical counterparts of  other transient systems \citep{Coe1985}. The extinction value of $A_v$ = 2.8 $\pm$ 0.3 mag. suggested the distance to be 4 kpc. From the Corbet diagram \citep{Corbet1984} for high magnetic field accreting pulsars, the orbital period is expected to be around 350 days. The known correlation between the orbital period of Be star binaries and $H_{\alpha}$ EW of the optical companion also indicates a similar large orbital period \citep{Reig1997} for 1A 1118--61. However, recently, detection of a 24-day binary period was reported by \cite{Staubert2010} using {\em RXTE}/PCA data.  

   {\em Einstein} and {\em EXOSAT} observations of this source were carried out in 1979 and 1985 respectively. During these observations the source was in a quiescent state and the luminosity calculated from the {\em EXOSAT}/ME observations was 0.5-3.0 $\times$10$^{34}$ ergs s$^{-1}$ at 3-7 kpc. This confirms that in the quiescent state, centrifugal inhibition of accretion was not complete. Three outbursts have so far been detected in this source. First outburst was in 1974 \citep{Maraschi1976}. The source had a second outburst that was first detected with Burst and Transient Source Experiment (BATSE) on the {\em Compton Gamma Ray Observatory} ({\em CGRO}) in 1991/1992. During this period the source had a peak flux of about 145 mCrab and a spin down rate of 0.016 s/day \citep{coe1994}. The source was also observed by the WATCH all sky monitor on {\em Granat} \citep{Lund1992}. 

1A 1118--616 was in  quiescence for $\sim$20 years and became highly active in 2009 January. The main outburst lasted only for about $\sim$20 days. This third outburst, observed by {\em Swift}/XRT in January 2009 revealed a pulsation period of 407.68 s \citep{Mangano2009a} indicating a spin-down in between the outbursts. About three weeks after the start of the outburst, the source was also observed with the {\em International Gamma-Ray Astrophysics Laboratory} ({\em INTEGRAL}/JEM-X/ISGRI) which detected a flaring activity after the main outburst \citep{Leyder2009}. Many observations of this source were carried out with {\em Rossi X-ray Timing Explorer}/Proportional Counter Array ({\em RXTE}/PCA) during this period, and the combined analysis of the {\em RXTE}-PCA and High Energy X-ray Timing Explorer (HEXTE) spectra revealed a cyclotron line absorption feature at 55 keV which gives a magnetic field strength of 4.8 $\times$ 10$^{12}$ G for the neutron star \citep{Doroshenko2010}. 

Most of the transient High Mass X-ray Binary (HMXB) pulsars are known to have a Be star companion. In Be/X-ray binaries, X-ray emission is thought to be due to the accretion of matter by the neutron star from the slow, dense, radial outflow of the Be star \citep{Negueruela1998}. These systems are observed to exhibit two different types of X-ray outbursts. One is short X-ray outbursts (Type-I outbursts) lasting for a few days ($L_x \leq 10^{36} - 10^{37}$ ergs s$^{-1}$) occuring in several successive binary orbits at orbital phase close to the time of periastron passage and other is giant X-ray outbursts (Type-II outbursts) lasting for several weeks ($L_x \geq 10^{37}$ ergs s$^{-1}$) and may start at any orbital phase. The current outburst in 1A 1118-61 appears to be a Type-II outburst but of shorter duration and smaller peak luminosity than the Type-II outbursts in most other Be/X-ray binary pulsars.

We have carried out a detailed timing and spectral analysis of the {\em RXTE}-PCA observation of this source during the 2009 January outburst, to detect any intensity or energy dependence of the pulse profile, aperiodic variabilities and also to find a suitable spectral model in the 3-30 keV band. One component of the aperiodic variabilities seen in X-ray binaries is the Quasi Periodic Oscillations (QPO), generally thought to be related to the innermost regions  of the accretion disk. Any inhomogeneous matter distribution or blobs of material in the inner disk may result in QPOs in the power spectrum. This can give useful information about the interaction between accretion disk and the central object at different intensity levels. HMXB pulsars show QPOs only at low frequency, i.e. in the range of 10 mHz upto about 1 Hz. Black hole X-ray binaries and low magnetic field neutron stars show QPOs over a wide range of frequency from a few Hz to a few hundred Hz.
Studying QPOs and their variations with energy and luminosity gives important clues about the mechanism of QPO production. In section 2 we describe the observations and the data used in the present work. In section 3 we present the pulsation analysis, the power density spectra, the pulse phase averaged and pulse phase resolved spectroscopy using the {\em RXTE}-PCA archival data followed by a discussion of the results in section 4.
  
\section{OBSERVATIONS AND DATA}

\begin{figure}
\centering
\includegraphics[width=8 cm, angle=-90]{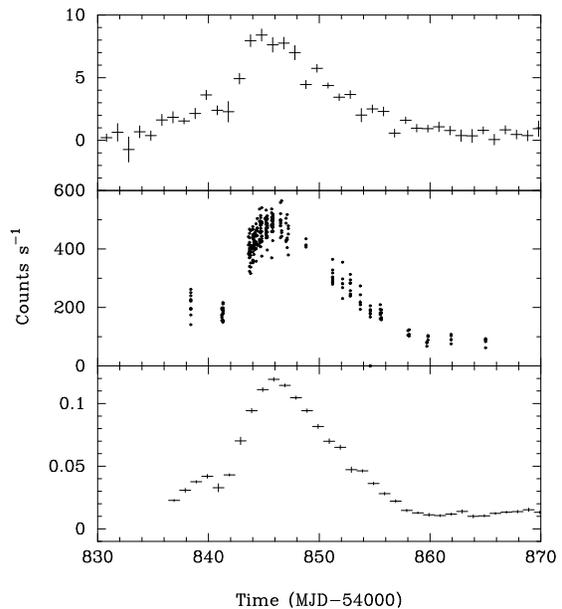}
\caption{The top panel shows the {\em RXTE}-ASM light curve of 1A 1118--61 in the 2-15 keV energy band with one day bin size, the middle panel shows the {\em RXTE}-PCU2 light curve with a bin size same as spin period and the bottom panel shows the 15-50 keV light curve from the {\em Swift}-BAT all sky monitor with a one day bin size.}
\label{asm-pcu}
\end{figure}

During the 2009 outburst of the pulsar 1A 1118--61, a series of observations were carried out with the {\em Rossi X-ray Timing Explorer} ({\em RXTE}). The {\em RXTE} consists of two non-imaging instruments: PCA \& the  HEXTE and an All Sky Monitor (ASM) that is sensitive to X-ray  photons between 1.5-12 keV. The data analyzed in this paper are from observations carried out using the PCA detectors. The PCA consists of five Xenon proportional counter units, sensitive in the energy range of 2-60 keV with a total effective area of 6500 cm$^{2}$.
A total of 144 ks of useful PCA data  was obtained from 26 pointings carried out during the outburst. For most of the pointings only one or two Proportional Counter Units (PCUs) were ON. The long term light curves of 1A 1118--61 are shown in Figure~\ref{asm-pcu} for different energy bands. The top panel shows the ASM light curve in 2-15 keV energy band with one day bin size, the middle panel shows the PCU2 light curve with a bin size same as spin period and the bottom panel shows the 15-50 keV light curve from the {\em Swift}-BAT all sky monitor with a one day bin size. For the timing analysis presented in the next section, we have used Standard1 mode data of the PCA having 0.125 s time resolution and Good Xenon mode data. We have used Standard2 mode data for spectral analysis and the spectral fitting was performed using {\em XSPEC} v12.6.0.


 \begin{figure}
\centering
\includegraphics[width=5.3 cm, angle=-90]{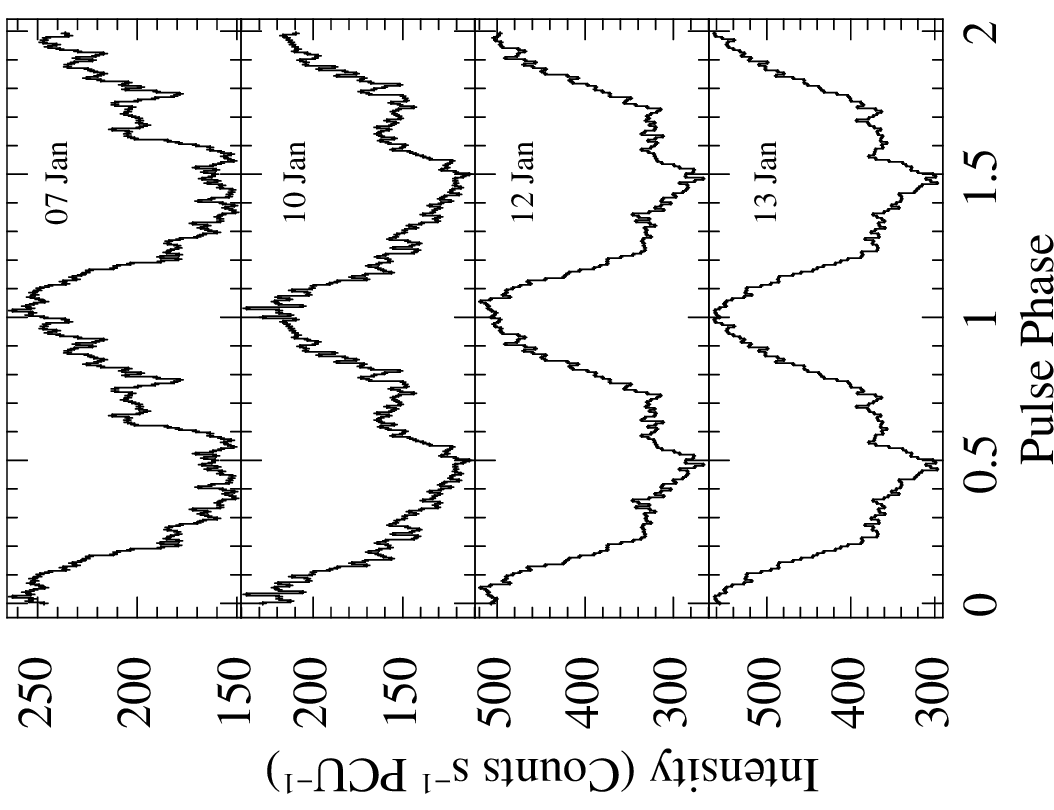}
\includegraphics[width=5.3 cm, angle=-90]{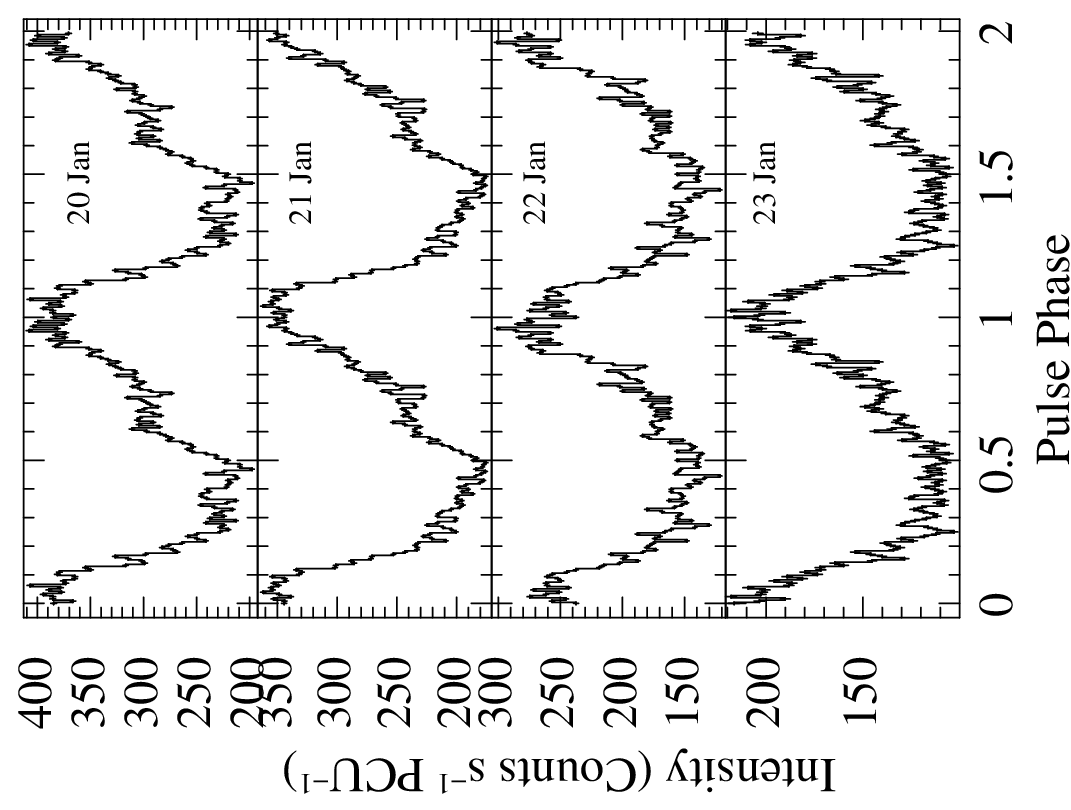}
\includegraphics[width=5.3 cm, angle=-90]{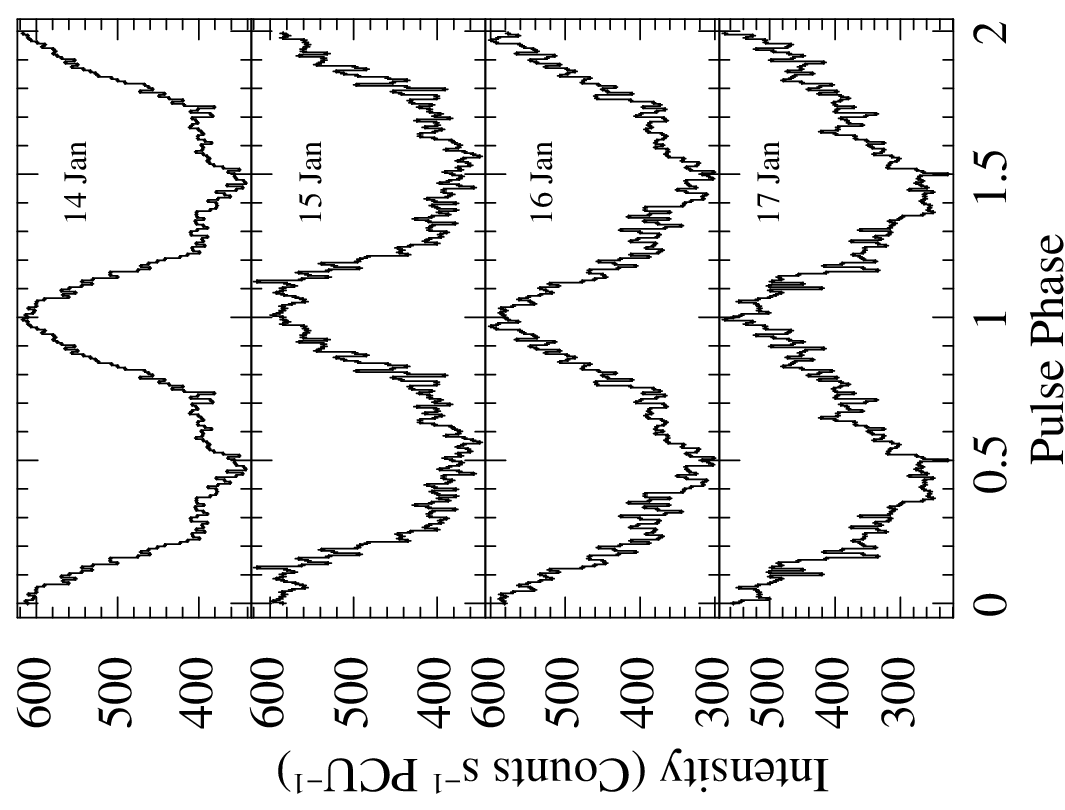}
\includegraphics[width=5.3 cm, angle=-90]{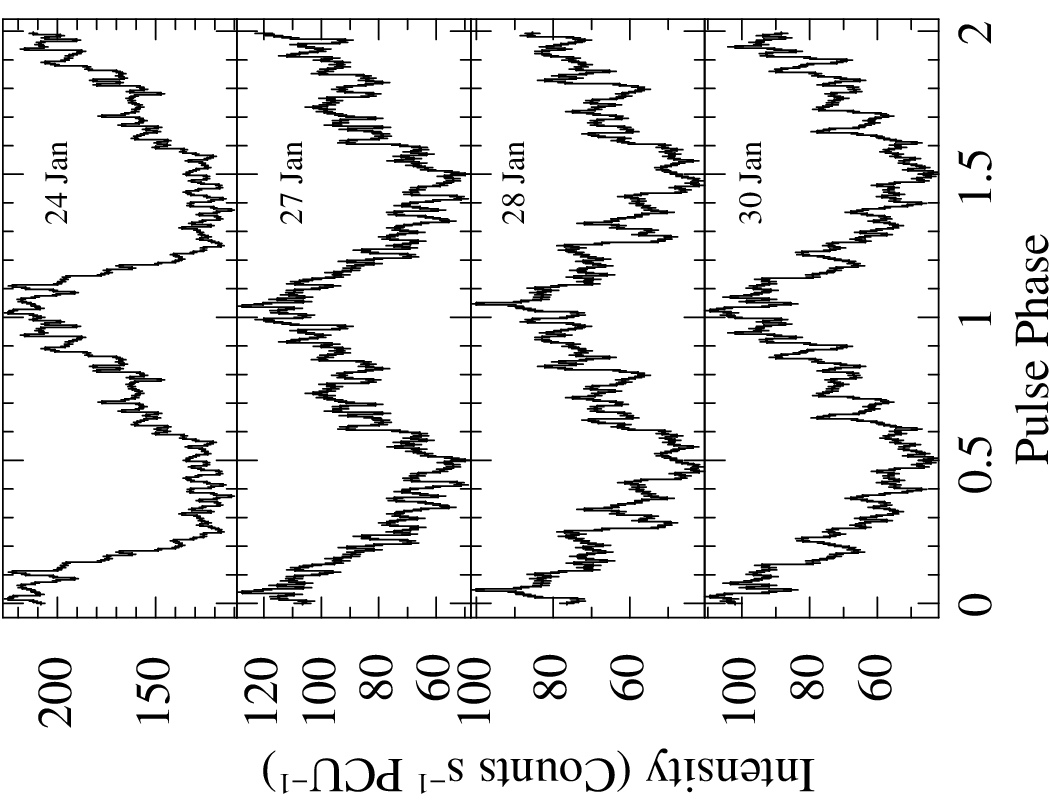}
\caption{The pulse profiles of 1A 1118--61, in the 2-60 keV band obtained on different days observed with {\em RXTE}-PCA during the outburst.}
\label{fig2}
\end{figure}

\section{Timing Analysis}
 \subsection{Pulse profiles}
 
We created 2-60 keV light curves of the source with a time resolution of 0.125 s from the Standard1 mode data. To remove data acquired during periods of Earth occultations and satellite unstable pointing, we have created Good Time Intervals (GTI) by screening the raw data with an offset of $<$ 0.02 degrees and elevation angle $>$ 10 degrees. Subsequently, the background count rate was subtracted from the light curves. The 2009 outburst lasted for $\sim$20 days starting around MJD 54834 and reached its maximum at around MJD 54845. The source count rate reached a peak value of about 600 c/s per PCU. In all but the last few PCA observations, the pulsations at around 408 s is clearly seen in the light curves. To measure the pulse period and its changes, we first carried out a barycenter correction and then applied the pulse folding and $\chi^{2}$ maximization technique. All the single data stretches were not long enough to measure the pulse period with high enough accuracy to investigate the pulse period evolution during this outburst. The most accurate pulse period measurement we obtained from any single observation is 407.58 s (MJD 54845). The pulse profile for each day were created by folding the light curve at the pulse period determined for that day and are shown in Figure~\ref{fig2}. The pulse profile showed significant evolution in shape during the outburst. At the beginning of the outburst, the pulse profile is complex with a smaller peak before the main peak. As the outburst progressed, the amplitude of the second peak decreased and on some days the pulse profile appears to have a main peak, a narrow minimum offset by 0.5 spin phase with respect to the main peak and two steps in between the main peak and the minimum, like a shoulder. At the end of the outburst, the overall shape of the pulse profile is a simple sinusoid, but with many narrow spikes or flares. 

\begin{figure}
\centering
\includegraphics[width=5.3 cm, angle=-90]{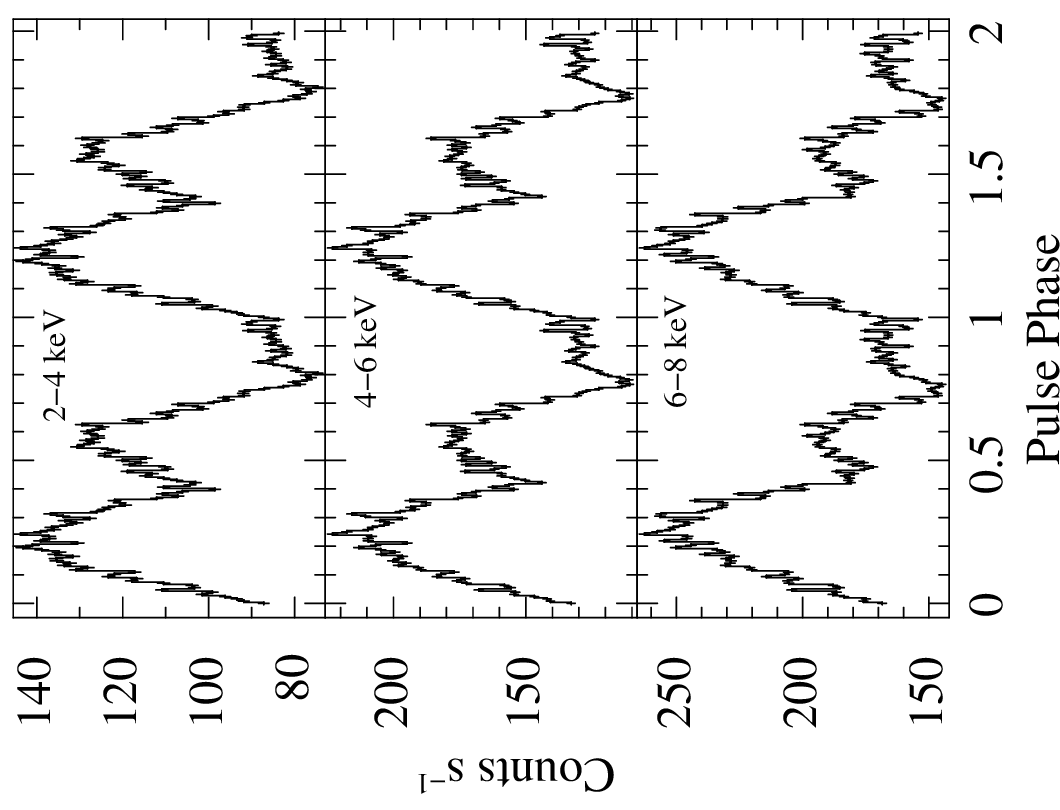}
\includegraphics[width=5.3 cm, angle=-90]{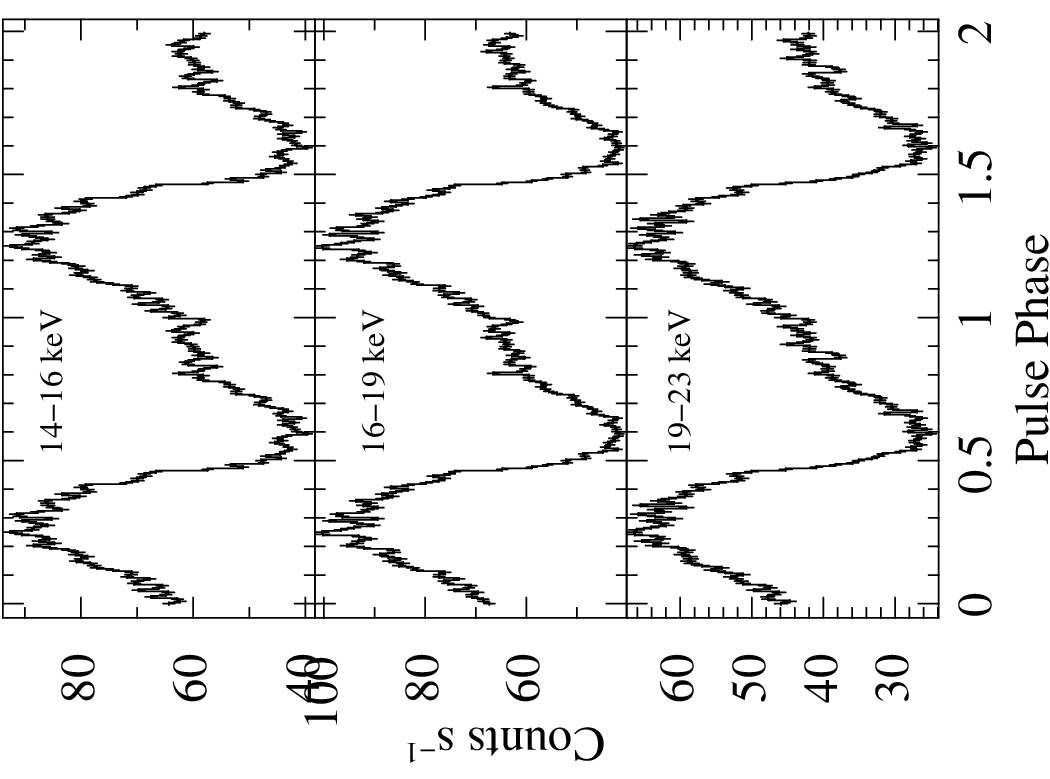}
\includegraphics[width=5.3 cm, angle=-90]{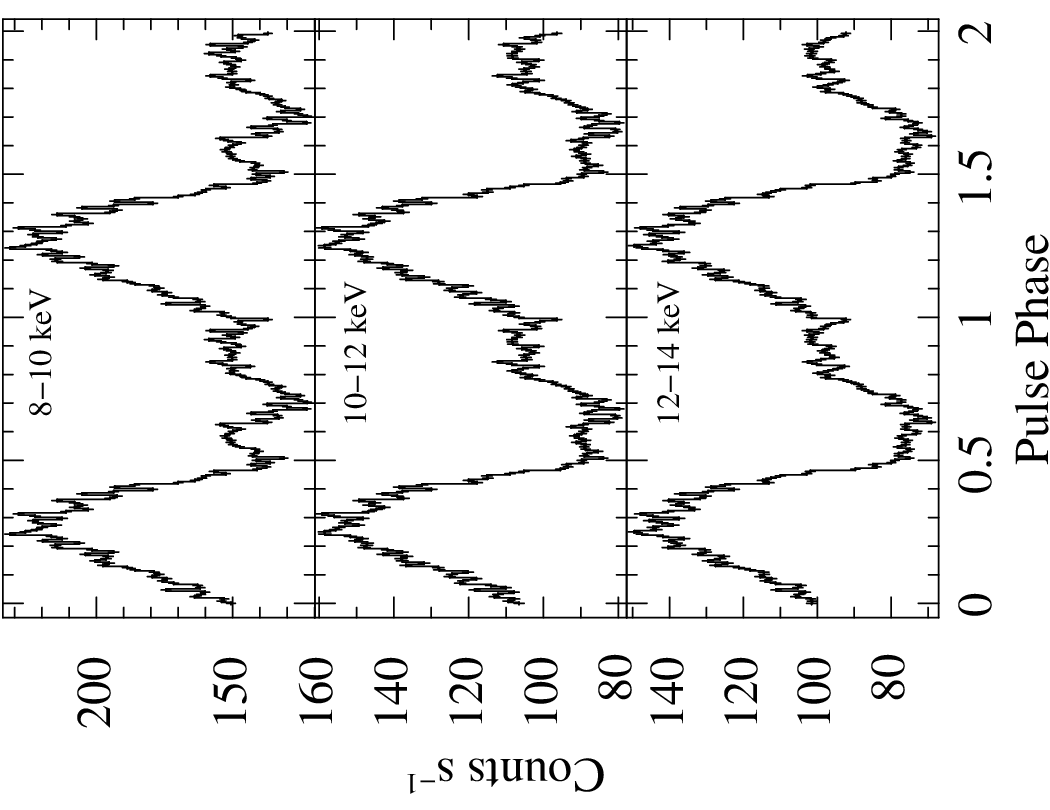}
\includegraphics[width=5.3 cm, angle=-90]{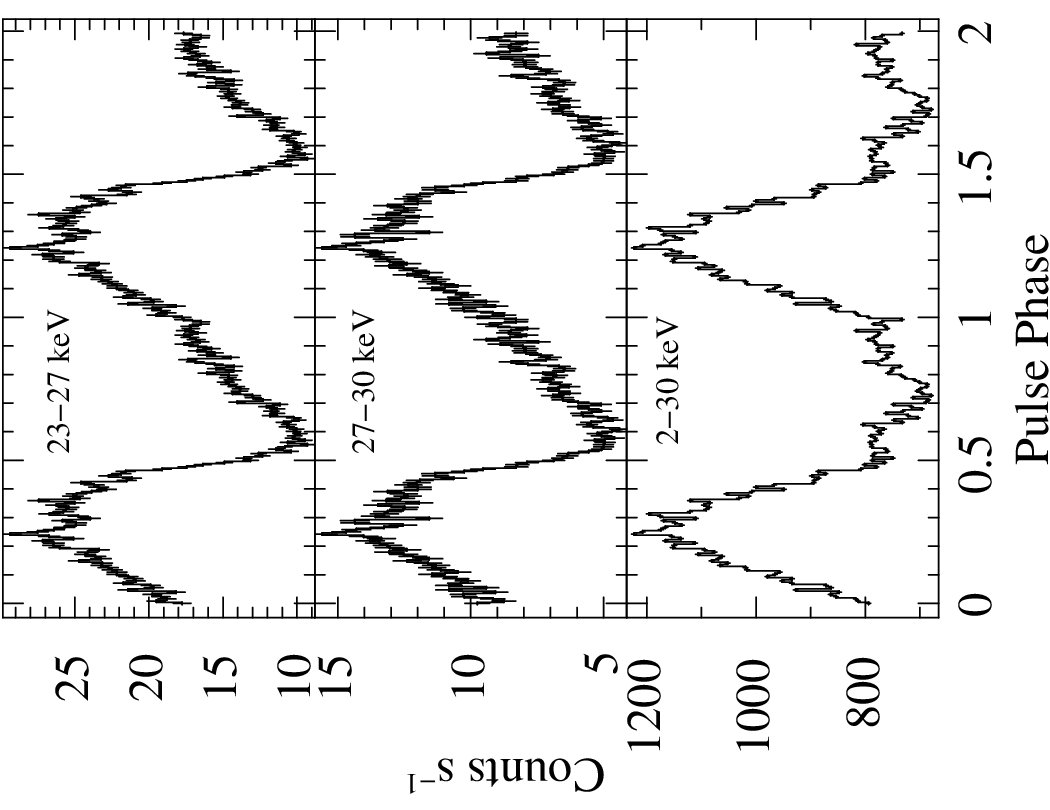}
\caption{Energy dependent pulse profiles of 1A 1118--61 in different energy bands observed with {\em RXTE} during the peak phase of outburst}
\label{fig3}
\end{figure}

   To check for any energy dependence of the pulse profiles of 1A 1118--61, we created light curves using Good Xenon mode data. We selected the observation (Obs ID-P94032/94032-04-02-07) carried out on Jan 14  near the peak of the outburst having the highest signal to noise ratio.
The profiles generated for different energy ranges are shown in Figure~\ref{fig3}. At low energies, the pulse profile has two peaks with the main peak leading by a phase of about 0.4, a dip after the smaller peak and a step before the main peak. At high energy, the pulse is made of a single asymmetric peak aligned with the main peak in the low energy pulse profile. The energy dependence of the pulse profile is seen to be very complex in 1A 1118--61. The smaller peak, seen quite clearly in low energy range, becomes weaker with increasing energy and above 10 keV, instead of a peak, the minimum of the pulse profile occurs at this phase. The phase range at which the minimum and the step occurs in the low energy band, grows in intensity with increasing energy and at high energy this becomes the leading part of the pulse peak. We investigated this complex energy dependence of the pulse profile by carrying out pulse phase resolved spectroscopy described later.

\subsection{Power Density Spectrum}

\begin{figure}
\centering 
\includegraphics[width=5.5 cm, angle=-90]{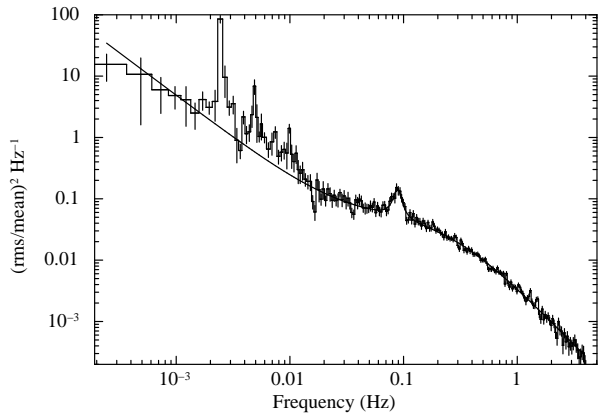}
\caption{A representative power spectrum of 1A 1118--61 showing a QPO feature at 0.09 Hz (MJD 54841-MJD 54854), spin frequency and its harmonics.}
\label{fig4}
\end{figure}

We created power density spectrum (PDS) from the 2-60 keV light curves of 1A 1118--61
using the FTOOL-{\em powspec}. The light curves were divided into stretches of length 4096 s
and the PDS obtained from each of these segments in one observation were averaged to produce
the final PDS. The expected white noise level was subtracted and the PDS was normalized such
that the integral gives the squared rms fractional variability. A peak at around $0.0025$ Hz
corresponding to the spin frequency and its harmonics are seen clearly in all the PDS except
in the last few observations. In addition to the main peak, a QPO feature is also seen at 0.09 Hz.
The continuum of the power spectrum was fitted with a model consisting of a power law component
and a Lorentzian and the QPO feature was fitted by adding another Lorentzian. The frequency
bins corresponding to the pulse frequency and its harmonics were avoided while fitting the continuum. 
The QPO feature was detected during the period 54841 to 54854 MJD. A representative power spectrum
with the QPO feature, is shown in Figure~\ref{fig4}.

\begin{figure}
\centering 
\includegraphics[width=6 cm, angle=-90]{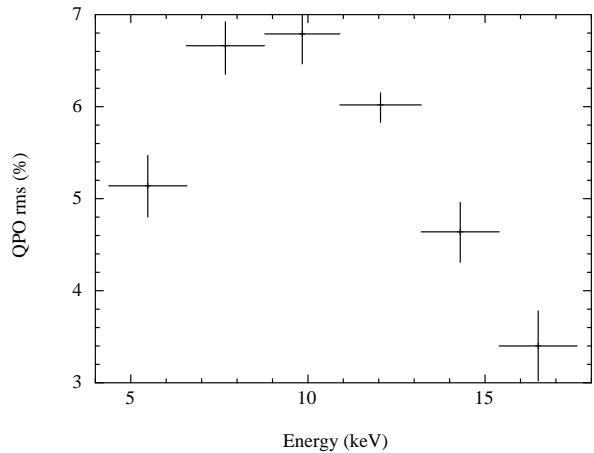}
\caption{RMS fluctuation in the 0.09 Hz QPO is shown here as a function of energy.}
\label{fig5}
\end{figure}

\begin{figure}
\centering 
 \includegraphics[width=6 cm, angle=-90]{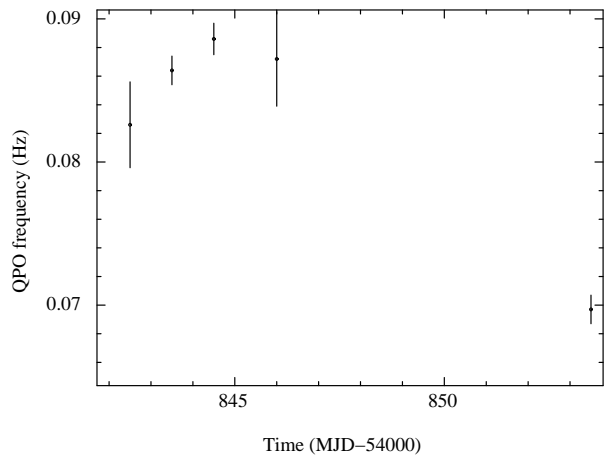}
\caption{Top panel shows the variation of the QPO frequency as a function of time and the bottom panel shows the energy averaged
rms value as a function of time.}
\label{fig6}
\end{figure}

The Quality factor ($\nu/FWHM$) of the QPO feature was about 5.4 and the detection significance of the QPO
feature, including data from all the days when the QPOs were detected is 7$\sigma$. The rms fractional variability
calculated from the background subtracted data was $5.2\%$. Using the event mode data, we have measured the
energy dependence of the QPO rms which is shown in Figure~\ref{fig5}. The rms value of the QPO is found
to increase up to 9 keV and then decrease. QPOs were detected on five different days and a plot of the QPO
frequency and rms fractional variability on these days is shown in Figure~\ref{fig6}. The QPO frequency decreased
from 0.09 Hz to 0.07 Hz over a span of 10 days along with the decay of the outburst.

\section{Spectral Analysis}

\subsection{Pulse Phase Averaged Spectroscopy}
 
To study the spectral variations during the outburst, we selected observations carried out on Jan 07, 14, 20, 21, 24 and Jan 30 covering the entire outburst. We used Standard2 mode data collected with PCU2 detector only to extract the source spectrum. The background spectrum was simulated using the {\em pcabackest} tool and appropriate background models for bright sources provided by the XTE guest observer facility (GOF) was used. We have used 0.5$\%$ systematic error to account for the calibration uncertainties while fitting the spectra. We first tried to fit the 3-30 keV spectrum with a model consisting of an absorbed power-law with a high energy cut off and a Gaussian for the iron line emission. This did not give a satisfactory fit with reduced $\chi^{2}$ in the range of 1.6 - 9.3. Addition of a black body component improved the fit, with reduced $\chi^{2}$ in the range of 1.0 - 2.0 for all six spectra. Temperature of the black body component was obtained as about 0.18 - 0.36 keV on different days, which is common for bright accretion powered pulsars (Paul et al. 2002). However, we found that with this model, the pulse phase resolved spectra could not be fitted for all phase bins. For some of the phase resolved spectra, it gives a rather high reduced $\chi^{2}$ of more than 8.0. If the complex energy dependence of the pulse profile is produced by different pulse phase dependence of two different spectral components, the two should dominate at different energies; one above and one below the energy at which the pulse shape changes, ie. $\sim$10 keV for 1A 1118--61 (Figure~\ref{fig3}). A blackbody component of temperature of a few hundred eV is therefore unlikely to produce the observed energy dependence of the pulse profile.

\begin{figure}
\centering
\includegraphics[height=4 cm, width=4 cm, angle=-90]{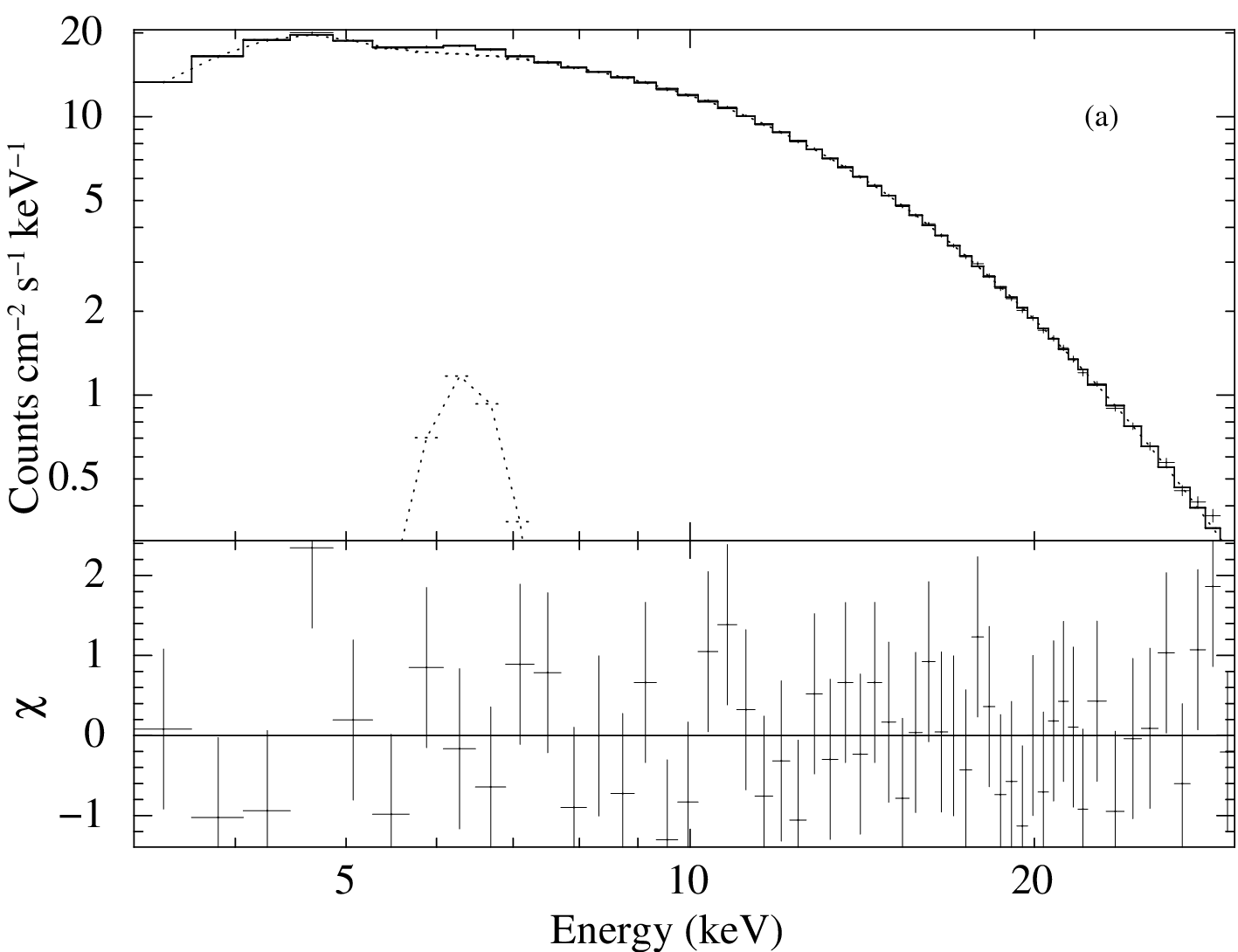}
\includegraphics[height=4 cm, width=4 cm, angle=-90]{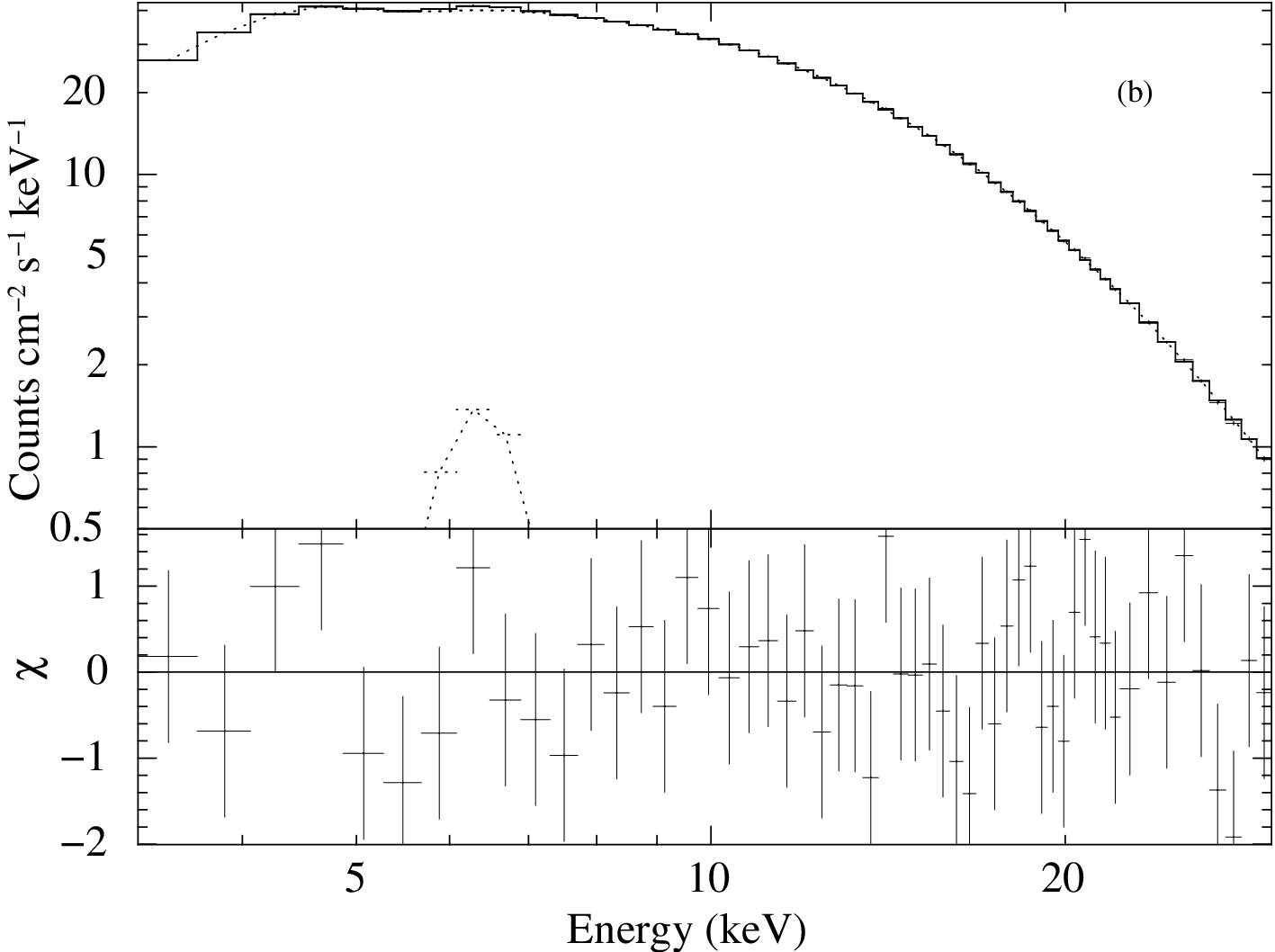}
\includegraphics[height=4 cm, width=4 cm, angle=-90]{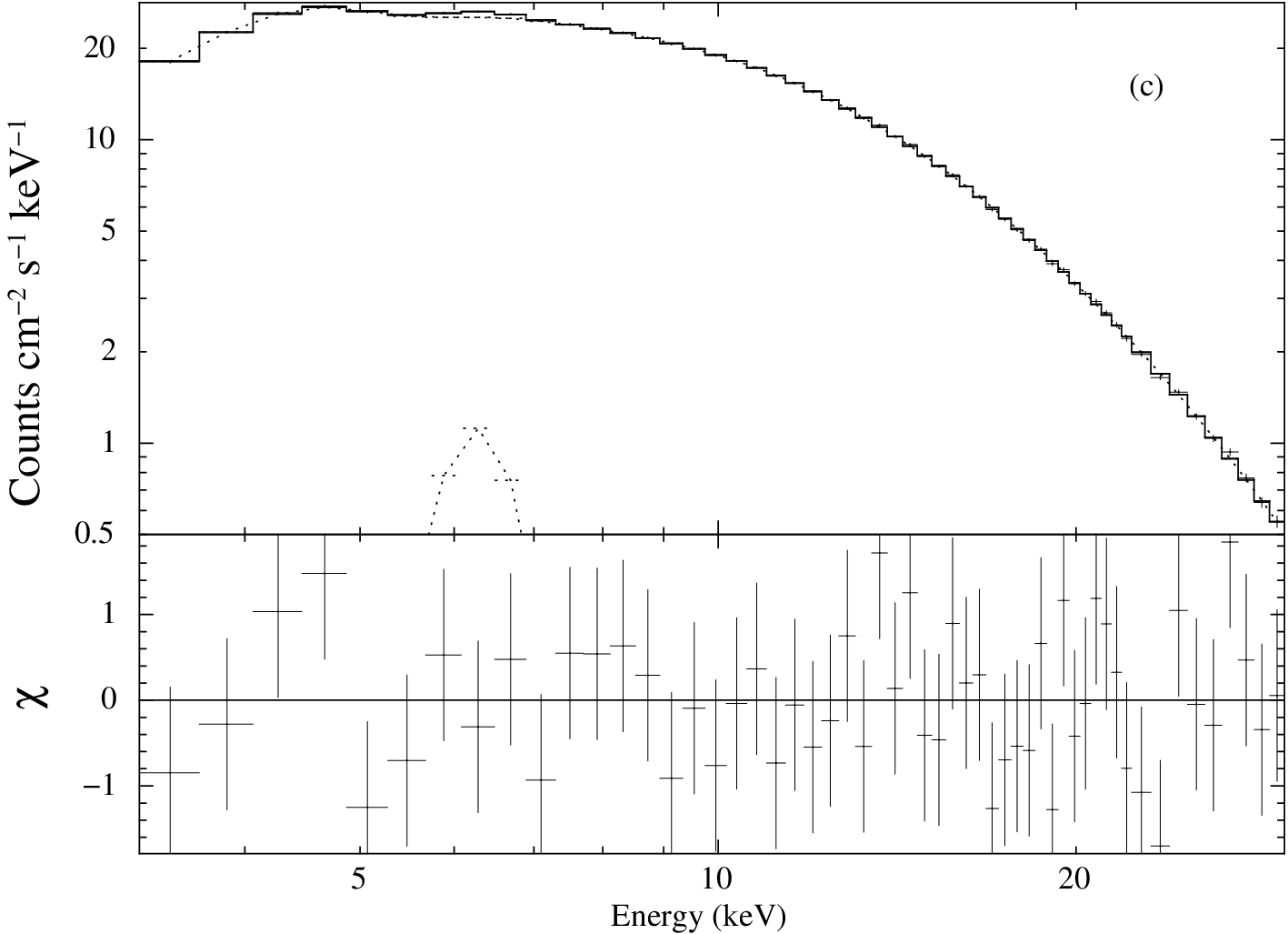}
\includegraphics[height=4 cm, width=4 cm, angle=-90]{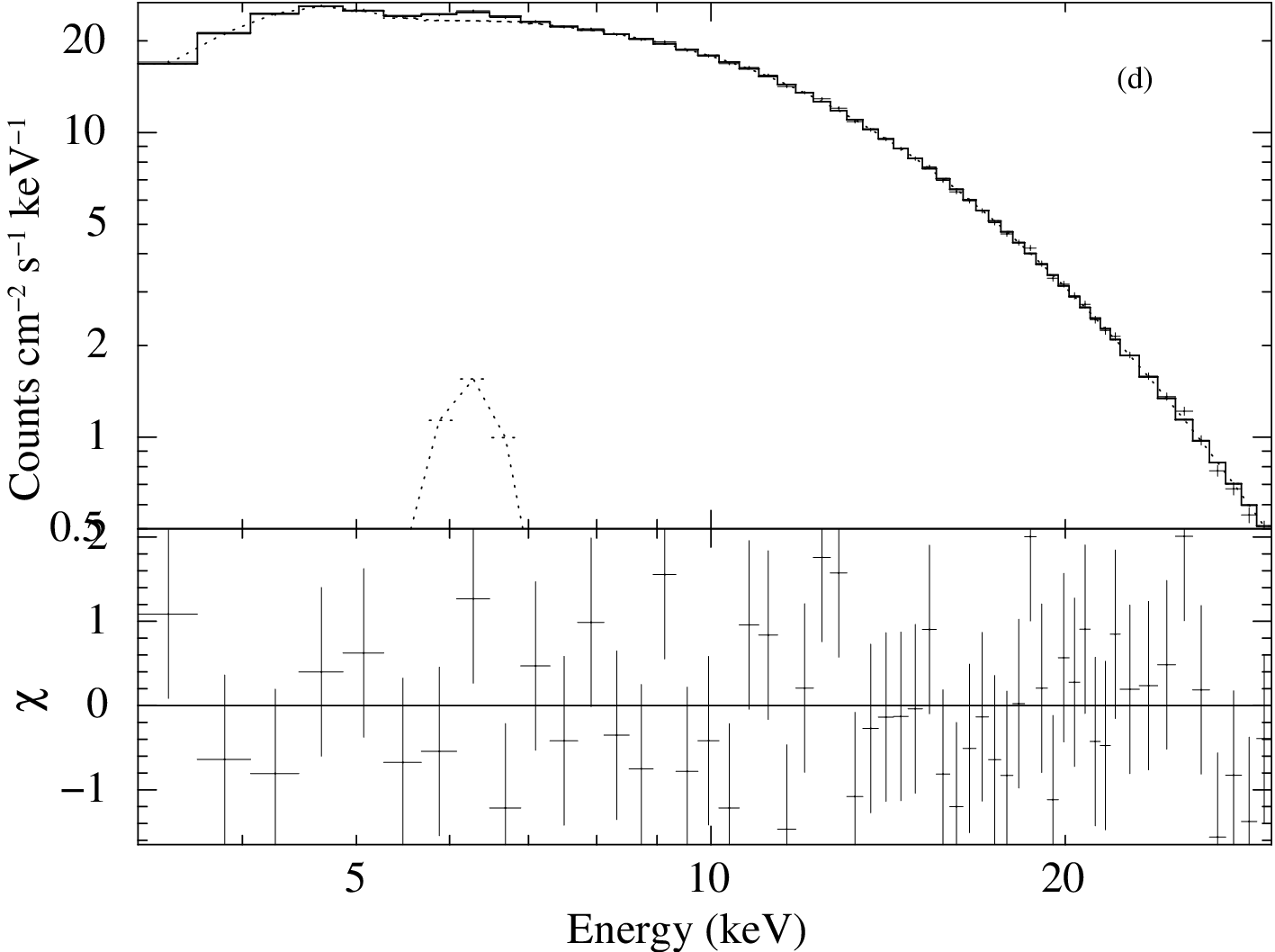}
\includegraphics[height=4 cm, width=4 cm, angle=-90]{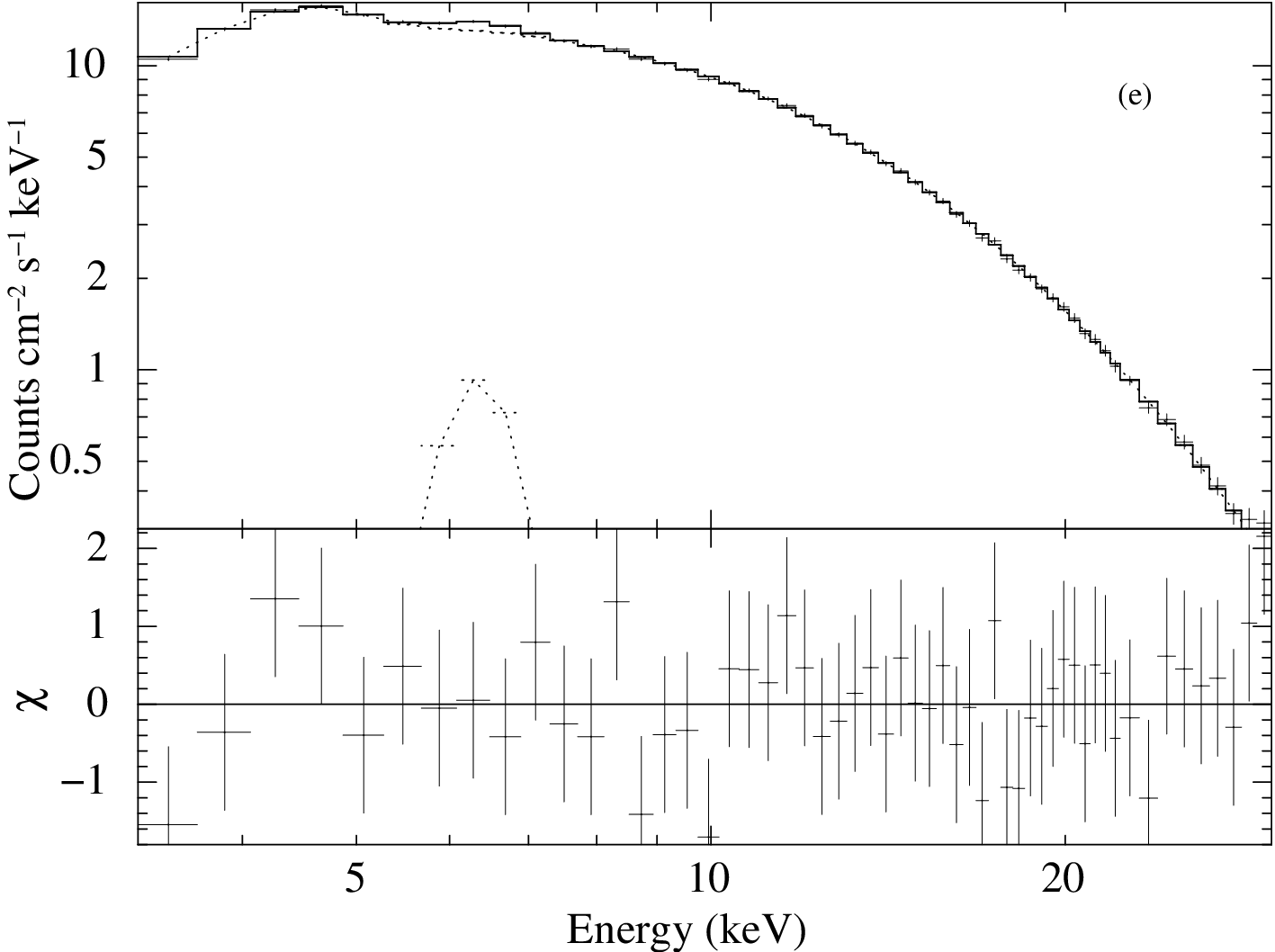}
\includegraphics[height=4 cm, width=4 cm, angle=-90]{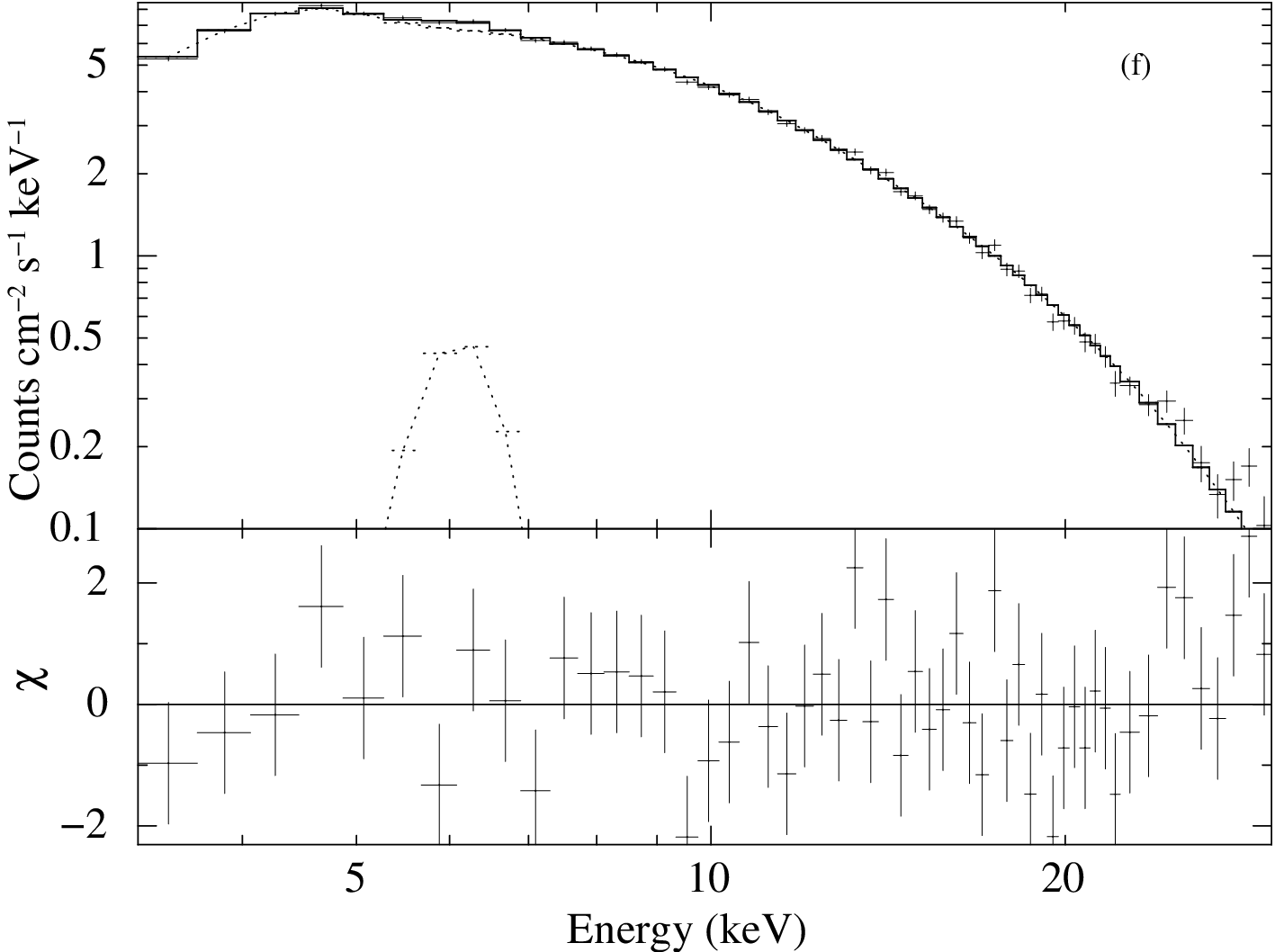}
\caption{3-30 keV energy spectra of 1A 1118--61 are shown here during different stages of the outburst along with the best fitted spectral model and the residuals; (a) the beginning of the outburst, (b) the peak, (c) \& (d) decay phases, (e) the end \& (f) the tail of the outburst}
\label{fig7}
\end{figure}

On the other hand, we find that a model consisting of partial covering of a power-law with high energy
cutoff and a Gaussian line emission describes all the pulse phase averaged spectra and the pulse phase
resolved spectra quite well. A partial covering model is known to fit a wide range of X-ray spectra
of HMXB pulsars \citep{Mukherjee2004}. In the present case, this model does not leave any soft excess
and thus does not require a black body component. The reduced $\chi^{2}$ is also smaller compared to 
other models. Spectral parameters obtained for 1A 1118--61 on the six different days are given in Table ~\ref{spec-para}.
The errors given here are for 1 $\sigma$ confidence level. The spectra, along with the best fitted model and
the corresponding residuals are shown in Figure ~\ref{fig7}.

\begin {table*}
\caption{Best fit Spectral Parameters of 1A 1118--61}
~\\
\begin {tabular}{|c|c|c|c|c|c|c|}
\hline
\hline
Parameter & 07Jan &14Jan&20Jan&21Jan &24Jan &30Jan\\
\hline
N$_{H1}$(10$^{22}$ cm$^{-2}$)$^a$&1.22 (fixed)&1.22 (fixed)& 1.22 (fixed)&1.22 (fixed)&1.22 (fixed) &1.22 (fixed)\\[6pt]

N$_{H2}$(10$^{22}$ cm$^{-2}$)&$284_{-48}^{+49}$&$216_{-19}^{+20}$ &$218_{-29}^{+28}$ &$371_{-130}^{+127}$ &$341_{-46}^{+48}$&$639_{-64}^{+93}$ \\[6pt]

{\em CvFract}&$0.15_{-0.04}^{+0.03}$&0.17$\pm 0.02$ &0.15$\pm 0.03$ &0.11$\pm 0.05$ & 0.26$\pm 0.05$&$0.54_{-0.04}^{+0.03}$\\[6pt]

P$_{Index}$&0.50$\pm 0.03$&0.30$\pm 0.02$&0.46$\pm 0.03$ &0.34$\pm 0.05$ &0.55$\pm 0.05$&$0.44_{-0.08}^{+0.05}$\\[6pt]

P$_{Norm}$&$0.054_{-0.002}^{+0.003}$&0.083$\pm 0.004$ &0.069$\pm 0.004$ &$0.053_{-0.002}^{+0.003}$ &0.053$\pm 0.002$&0.037$\pm 0.03$\\[6pt]

E$_{cut}$ (keV)&$4.96_{-0.17}^{+0.20}$&$5.88_{-0.19}^{+0.17}$ &$5.84_{-0.28}^{+0.22}$ &$4.88_{-0.28}^{+0.32}$  &$4.76_{-0.23}^{+0.25}$&$5.01_{-0.23}^{+0.20}$\\[6pt]

E$_{fold}$ (keV)&$12.62_{-0.59}^{+0.65}$&$12.38_{-0.24}^{+0.28}$&14.08$\pm 0.48$&$13.12_{-0.92}^{+0.74}$ &$13.22_{-1.06}^{+1.08}$&$8.35_{-0.62}^{+0.60}$ \\[6pt]

E$_{Fe}$ (keV)&6.41$\pm 0.06$&$6.42_{-0.08}^{+0.07}$&$6.33_{-0.09}^{+0.08}$ & 6.30$\pm 0.06$  &6.41$\pm 0.08$&6.17$\pm 0.10$\\[6pt]

Fe$_{EqWidth}$ (eV)&84&41 &52 &78 &86&86\\[6pt]

Fe$_{Norm}$$^b$&1.34&1.57&1.27 &1.76 &1.05&0.57\\[6pt]

Total Flux (3-30 keV)$^c$&3.25&8.91&5.39&5.03&2.62&1.11\\[6pt]

Red-$\chi^{2}$/d.o.f&0.76/47&0.78/47&0.77/47&0.96/47&0.67/47&1.36/47\\[6pt]
\hline
\end{tabular}

$ ^a ${The minimum of N$_{H1}$ was fixed to the Galactic hydrogen column density of 1.22 $\times$ 10$^{22}$ atoms cm$^{-2}$}\\ towards this source and was allowed to vary only to the higher side.\\ 
$ ^b ${$10^{-3}$ photons cm $^{-2}$ s$^{-1}$}\\
$ ^c ${$10^{-9}$ ergs cm$^{-2}$ s$^{-1}$ for 3-30 keV}
\label{spec-para}
\end{table*}

 \subsection{Pulse Phase Resolved Spectroscopy}

\begin{figure}
\centering
\includegraphics[width=8.5 cm, angle=-90]{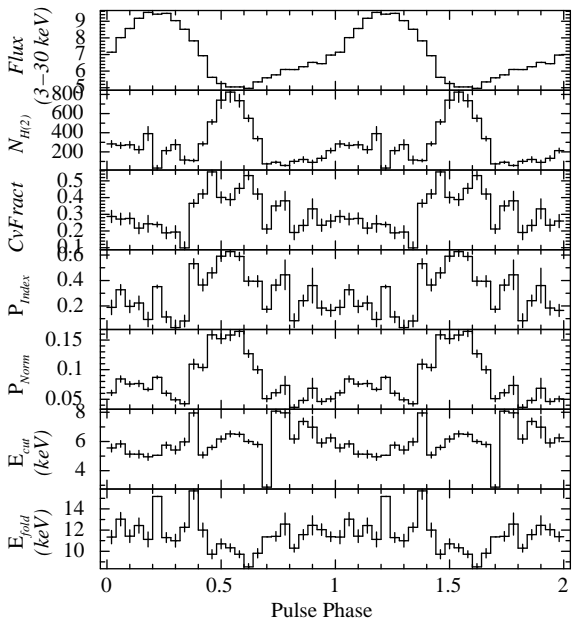}
\caption{Variation of spectral parameters with phase is shown. Flux (3-30 keV) in units of 10$^{-9}$ ergs cm$^{-2}$ s$^{-1}$ and N$_{H (2)}$  in units of 10$^{22}$ atoms cm$^{-2}$}
         
\label{fig8}
\end{figure}

The pulse profiles shown in Figure~\ref{fig3} exhibit strong energy dependence. As we have argued, a low temperature black body component cannot explain the features above 3 keV. So to investigate this in detail, we have performed pulse phase resolved spectroscopy of an observation made near the peak of the outburst (MJD 54845) which has the highest signal to noise ratio among all the {\em RXTE}-PCA observations. Energy spectra were extracted in 25 phase ranges and appropriate background subtraction was done. The partial covering model gave acceptable fit for the spectra in all pulse phase bins. The iron line parameters and the first hydrogen column density were fixed to the phase averaged value and the rest of the parameters were allowed to vary. We could not constrain the $E_{fold}$ and $E_{cut}$ values for phases 0.22 and 0.70. So we freeze the $E_{cut}$ and $E_{fold}$ parameter values to the phase averaged values for these two phases. The variation of the free parameters with pulse phase is shown in Figure~\ref{fig8}. We found significant variation in all the spectral parameters. The second column density and the covering fraction varied by a large factor. However, we note that variation of two of the parameters, the power-law normalization and the photon index are correlated and may not be entirely true.

\section{Discussion}

\subsection{Quasi Periodic Oscillations}

QPOs have so far been detected in 19 accretion powered high magnetic field pulsars which include mostly HMXBs and a few LMXBs, in both transient and persistent sources. In Table 2 we have listed the sources, the spin frequency $\nu_{s}$, the QPO frequency $\nu_{QPO}$ and its range, and the ratio of the two.

\begin {table*}
 \caption{List of QPO sources}
~\\
\begin {tabular}{|c|c|c|c|c|c|}
 \hline
Source&Type&$\nu_{s}$&$\nu_{QPO}$&$\nu_{QPO}$/$\nu_{s}$&Reference\footnotemark{}\\
    &   &(mHz)&(mHz)&&\\
\hline
Transient pulsars& & & &\\
\hline

KS 1947+300&HMXB/Be&53&20&0.38&1\\
SAX J2103.5+4545&HMXB/?&2.79&44&15.77&2\\
A0535+26&HMXB/Be&9.7&50&5.15&3\\
V0332+53&HMXB/Be&229&51&0.223&4\\
4U 0115+63&HMXB/Be&277&62&0.224&5\\
1A 1118--61&HMXB/Be&2.5&92&36.8&This work\\
XTE J1858+034&HMXB/Be&4.53&110&24.3&6\\
4U 1901+03&HMXB/?&361.9&130&0.359&7\\
EXO 2030+375&HMXB/Be&24&200&8.33&8\\
SWIFT J1626.6-5156 &HMXB/Be&65&1000&15.38&9\\
XTE J0111.2-7317&HMXB/B0.5-B1Ve&32&1270&39.68&10\\
GRO J1744-28&LMXB&2100&20000&9.52&11\\
\hline
Persistent pulsars& & & &\\
\hline
SMC X-1&HMXB/B0&1410&10&0.0071&12\\
Her X-1&LMXB&806&13&0.016&13\\
LMC X-4&HMXB/O-type&74&0.65-20&0.0087-0.27&14\\
Cen X-3&HMXB/O-type&207&35&0.17&15,16\\
4U 1626-67&LMXB&130&48&0.37&17,18\\
X Per&HMXB/Be&1.2&54&45&19\\ 
4U 1907+09&HMXB/OB&2.27&69&30.4&20,21\\
\hline
\end{tabular}
\vskip 1cm
\footnotetext{}{References:
(1) \cite{James2010};
(2)\cite{Inam2004};
(3)\cite{Finger1996};
(4) \cite{Takeshima1994};
(5) \cite{Soong&Swank1989};
(6) \cite{Paul&Rao1998};
(7) \cite{James2011};
(8) \cite{Angelini1989};
(9) \cite{Reig2008};
(10) \cite{Kaur2007};
(11) \cite{Zhang1996};
(12) \cite{Angelini1991};
(13) \cite{Moon2001b};
(14) \cite{Moon2001a};
(15) \cite{Takeshima1991};
(16) \cite{Raichur2008};
(17) \cite{Shinoda1990};
(18) \cite{Kaur2008};
(19) \cite{Takeshima1997};
(20) \cite{Zand1998}; 
(21) \cite{Mukerjee2001} }
\end{table*}

The most commonly used models for explaining the QPO mechanism are Keplerian frequency model (KFM), beat frequency model (BFM) and accretion flow instabilities. In the KFM, the QPOs arise due to inhomogeneities at the inner edge of the accretion disk modulating the light curve at the Keplerian frequency. In the BFM, the accretion flow onto the neutron star is modulated at the beat frequency between the Keplerian frequency of the inner edge of the accretion disk and the spin frequency $\nu_{QPO}= \nu_{k} -\nu_{s}$ \citep{Shabazaki&Lamb1987}. The third model applies only to the sources that have luminosities close to the Eddington limit \citep{Fortner1989}.

The occurrence of QPOs in accretion powered pulsars is quite complex. In some sources like 4U 1626--67, QPOs are detected most of the time while in some others like Cen X-3, QPOs are rare.  Recently, in two transient pulsars (KS 1947+300 and 4U 1901+03) QPOs were observed only at the end of the outbursts when the source intensity had fallen to a few percent of the peak of the outbursts \citep{James2010, James2011}. Some transient pulsars like XTE J1858+034, showed QPOs in all observations while some other transient pulsars like EXO 2030+375, showed QPOs in only some of the outbursts. In the 405 s recurrent transient pulsar 1A 1118--61, QPOs are observed in the range of 0.07 - 0.09 Hz in most of the observations near the peak of the outburst.
The frequency evolution of the QPO in 1A 1118--61 during the outburst and presence of QPOs during most of the outburst makes this source similar to the transient XTE J1858+034 \citep{Mukherjee2006}. In 1A 1118--61, the QPO frequency is higher than the spin frequency of the pulsar and can be explained in either KFM or BFM and since the QPO frequency is a few hundred times larger than the spin frequency, the radius at which the QPOs are generated is quite similar in the two models. We have,

\begin{equation}
R_{qpo} = \left({GM}\over{4 \pi^{2} \nu^{2}_{qpo}}\right)^{1/3}
\end{equation}

For an assumed neutron star mass of $1.4 M_{\odot}$, $R_{qpo}$ =  9$\times$ 10$^{3}$ km.

The average 3-30 keV X-ray flux during the period of QPO detection is 1.95$\times$10$^{-9}$ erg cm$^{-2}$ sec$^{-1}$,
which for a distance of 4 kpc corresponds to an X-ray luminosity of $0.37\times10^{37}$ erg sec$^{-1}$.
The radius of the inner accretion disk can be expressed in terms of the luminosity and magnetic moment
 as \cite{Frank1992}
\begin{equation}
R_{m}= 3\times10^{8}L^{-2/7}_{37}\mu^{4/7}_{30}
\end{equation}
where $L_{37}$ is the X-ray luminosity in units of 10$^{37}$ erg s$^{-1}$ and $\mu_{30}$ is the magnetic moment in units of 
10$^{30}$ cm$^{3}$ Gauss. Equating $R_{qpo}$ with $R_{m}$, we determine a magnetic moment of $\mu_{30}$ = 3.38 $\times$10$^{30}$ cm$^{3}$ G, which for a neutron star canonical radius of 10 km corresponds to a surface magnetic field strength of 3.38 $\times$10$^{12}$ G. The magnetic field strength of the neutron star derived using the QPO frequency and the X-ray luminosity is in excellent agreement with the strength of the magnetic field obtained from the cyclotron line absorption feature (4.8 $\times$10$^{12}$ G, Doroshenko et al. 2010).

\subsection{Pulse profile evolution and energy dependence}

Accretion powered X-ray pulsars are known to show interesting intensity dependence of the pulse profile, both in transient and persistent sources \citep{Nagase1989, Raichur2010}. A most remarkable example of pulse profile evolution was investigated in EXO 2030+375 with EXOSAT observations \citep{Parmar1989} . Evolution of the pulse profile during the outbursts indicate changes in the accretion flow from the inner accretion disc to the neutron star, especially changes in the mass accretion rate. The X-ray beaming pattern may also change during the outburst depending on the density, dimensions and structure of the accretion column. The 2-60 keV pulse profiles of 1A 1118--61, presented here also show a complex profile, and profile evolution, as the relatively short outburst reached its peak and then decayed. The pulse profiles shown in Figure 2 are clearly intensity dependent, but it is not a simple function of intensity. For example, for similar count rates per PCU, the pulse profile is quite different during the rise of the outburst (7-10 January 2009, Figure 2) and during the decay (22-23 January 2009, Figure 2).

Most accretion powered pulsars also show strong energy dependence of the pulse profile. In 1A 1118--61, the position of the main peak is at the same phase in all energies in the 2-30 keV band, but several other features including a second peak at low energy, a narrow dip in the low energy and a leading edge of the main peak at high energy are energy dependent features clearly seen in energy resolved pulse profiles (Figure 3). An intriguing aspect is that, the pulse shape change appears at around 10 keV. Strong energy dependence of pulse profiles - in other words strong pulse phase dependence of the energy spectrum - are known to be present also in magnetars \citep {Enoto2010} and multi component emission models are invoked there to describe the same. However, we find that multi component spectral model for accretion powered pulsars, including a low temperature black body cannot explain the change in pulse profile at around 10 keV. For pulse phase averaged and pulse phase resolved spectroscopy, we have therefore used a model that can fit features in the medium energy band.

\subsection{Pulse phase averaged and pulse phase resolved spectra}

The continuum spectra of accreting X-ray pulsars are often best described by a power law with a high-energy cutoff and interstellar absorption. And in most cases, the presence of a Gaussian component, for iron line fluorescent emission is also evident. Cyclotron resonance absorption features have also been detected in about 20 bright pulsars, usually at energies above 10 keV. In 1A 1118-61, Doroshenko et al. (2010) detected a cyclotron absorption feature at $\sim$55 keV using the combined PCA and HEXTE data in the 4-120 keV band. Considering the complex energy dependence of the pulse profile and from
pulse phase resolved spectroscopy we find that a partial covering power-law model describes the data very well. With a partial covering model we don't see a feature at 8 keV. The 8 keV feature mentioned in Doroshenko et al. (2010) and several other contemporary papers probably has same reason that a very simple, single continuum model is used while the sources have complex absorption. If the 8 keV feature is an instrument artifact, it should have been seen
in all kinds of sources. However, such a feature has not been reported in the X-ray spectrum of black hole binaries and is also not seen in the spectrum of the Crab Nebula \citep{Kirsch2005,Weisskopf2010}. We would also like to note here that at higher energy, the partial covering absorption model is the same as a simple power-law model, and thus the cyclotron absorption feature reported by Doroshenko et al. (2010) is not in question here. Lin et al. (2010) fitted the 0.5-10.0 keV energy spectrum of this source obtained with the {\em Swift}-XRT and found that in this limited energy band, the spectra are fitted best with two black body components. However, since the source shows strong hard X-ray emission (as detected with {\em Swift}-BAT, Figure 1) and the two blackbody model cannot produce the hard X-ray photons, this model is inappropriate for this pulsar as well as for any hard X-ray pulsar. 

In the partial covering absorption model, a part of the continuum source is obscured, resulting in a harder spectrum \citep{Wang2010}. If the absorbing component is in the form of an accretion stream or is a part of the accretion column, it can be phase locked with the neutron star, resulting into a phase dependent column density and covering fraction. The pulse phase resolved spectroscopy reported here shows a strong modulation of the partial covering fraction as well as the column density supporting such a scenario for this pulsar. We also notice a systematic variation of the cutoff energy value and the e-folding energy at the pulse phases with highest column density of the partial covering material. A similar energy dependence of the pulse profile and pulse phase dependence of the spectral parameters have recently been found in another accretion powered transient pulsar GRO J1008-57 \citep{Naik2010} in the broad band data obtained with the {\em Suzaku} X-ray observatory.

\section{Acknowledgments} 

We thank an anonymous referee whose suggestions helped us to improve the
content of the paper. This research has made use of data obtained
through the High Energy Astrophysics Science Archive Research Center
Online Service, provided by the NASA/Goddard Space Flight Center.

The QPO discovery mentioned in this paper was presented at the
38th COSPAR Scientific Assembly, July 2010, Bremen, Germany. After
initial submission of this paper, discovery of the same has also been
reported by Nespoli $\&$ Reig (2011, A$\&$A, 526,7).

\def\etal{{\it et~al.\ }}
\def\apj{{Astroph.\@ J.\ }}
\def\apjl{{Astroph.\@ J. \@ Lett. }}
\def\araa{{Ann. \@ Rev. \@ Astron. \@ Astroph.\ }}
\def\mn{{Mon.\@ Not.\@ Roy.\@ Ast.\@ Soc.\ }}
\def\aap{{Astron.\@ Astrophys.\ }}
\def\aj{{Astron.\@ J.\ }}
\def\prl{{Phys.\@ Rev.\@ Lett.\ }}
\def\pd{{Phys.\@ Rev.\@ D\ }}
\def\nucp{{Nucl.\@ Phys.\ }}
\def\nat{{Nature\ }}
\def\sci{{Science\ }}
\def\plb {{Phys.\@ Lett.\@ B\ }}
\def \jetpl {JETP Lett.}

\end{document}